%% file: main.tex
\documentclass[a4paper]{article}

\usepackage{spconf,amsmath,graphicx}
\usepackage{amsmath,amsfonts,bm}
\usepackage{amsmath,graphicx}
\usepackage{amssymb}
\usepackage{multirow}
\usepackage{xcolor}
\usepackage{tabularx}
\usepackage[skip=0.5\baselineskip]{caption}
\usepackage{enumitem}
\usepackage{booktabs}
\usepackage{listings}
\usepackage{setspace} 
\usepackage{url}

\input{math_commands.tex}

\ninept
\title{A Closer Look at Audio-Visual \\ Multi-Person Speech Recognition and Active Speaker Selection}
\name{Otavio Braga\textsuperscript{1}\thanks{\textsuperscript{1}obraga@google.com}, Olivier Siohan}
\address{Google, Inc.}

\begin{document}

\maketitle
\begin{abstract}

Audio-visual automatic speech recognition is a promising approach to robust ASR under noisy conditions. However, up until recently it had been traditionally studied in isolation assuming the video of a single speaking face matches the audio, and selecting the active speaker at inference time when multiple people are on screen was put aside as a separate problem. As an alternative, recent work has proposed to address the two problems simultaneously with an attention mechanism, baking the speaker selection problem directly into a fully differentiable model. One interesting finding was that the attention indirectly learns the association between the audio and the speaking face even though this correspondence is never explicitly provided at training time. In the present work we further investigate this connection and examine the interplay between the two problems. With experiments involving over 50 thousand hours of public YouTube videos as training data, we first evaluate the accuracy of the attention layer on an active speaker selection task. Secondly, we show under closer scrutiny that an end-to-end model performs at least as well as a considerably larger two-step system that utilizes a hard decision boundary under various noise conditions and number of parallel face tracks.

\end{abstract}

\section{Introduction}

  Up until recently, audio-visual ASR had been studied only under the ideal scenario where the video of a single speaking face matches the audio \cite{Makino19,Afouras_2018,Chung_2017,Potamianos2003,Saenko2006,Harte2015}. However, at inference time when multiple people are simultaneously visible on screen we need to decide which face to feed to the model, and this issue of active speaker selection has traditionally been considered a separate problem~\cite{ChungSyncNet_2017,ChungSyncnet2_2019}.
  
  As an alternative, it has been recently shown in~\cite{Braga20} that by using an attention mechanism we can address the two problems simultaneously, baking the speaker selection problem directly into a fully differentiable model that can be trained with backpropagation. Under this scenario, at inference time, all available faces are provided to the A/V ASR model, along with the audio. One interesting finding of this proposal was that the attention appears to indirectly learn the association between the audio and the speaking face even though this correspondence is never explicitly provided at training time.
  
  The purpose of the present work is to further investigate this finding and the interplay between the two problems. Specifically, we take a closer look at the attention layer and evaluate its accuracy on an active speaker selection task on the frame level. Secondly, we measure under stronger scrutiny the effectiveness of the proposed end-to-end approach versus a two-step system connected with a hard decision boundary. For the baseline, we couple a separately trained active speaker selection module with a state-of-the-art audio-visual ASR model that can handle a single face track. With experiments involving over 50 thousand hours of YouTube videos as training data, we show that the end-to-end model performs at least as well as the considerably bigger two-step system under various noise conditions and number of parallel face tracks, while still showing benefits over an audio-only model. This further solidifies the proposed end-to-end approach as a strong alternative to multi-person A/V ASR.
  
  It should be noted that the technology developed while working on this project abides by Google AI Principles \cite{GoogleAIPrinciples}\cite{Makino19}.

\vspace{-2mm}

\section{Models}

\subsection{Acoustic and Visual Frontends}

\noindent
{\bf Acoustic Features.} We employ mel filterbank features for both the ASR and speaker selection tasks. The 16kHz-sampled input audio is framed with 25ms windows smoothed with the Hann window function, with steps of 10ms between consecutive frames. We compute energies in 80 mel filter bank channels at each frame, compressing their range with a $\log$ function. We then fold every 3 consecutive feature vectors together, yielding a 240 dimensional feature vector every 30ms, which corresponds to acoustic features at about 33.3Hz. We denote the input acoustic features tensor by $\tA \in \R^{B\times T \times D_A}$, where $B$ is the batch size, $T$ is the number of time steps and $D_A$ ($=240$) the dimension of the acoustic features.

\

\noindent
{\bf Audio and Video Synchronization.} The videos in our training set have frame rates ranging from around 23 to 30 fps, so in order to make the input uniform we synchronize the videos with the acoustic features by resampling the video with nearest neighbor interpolation in time at the acoustic features sample rate (33.3Hz). Since video frames are synchronized with the acoustic features, {\it frames} will be used interchangeably for video and speech frames in this paper. In the spatial dimension, we crop the full face tracks around the mouth region to generate images of resolution $128\times128$, with RGB channels normalized between $-1$ and $1$.

\

\noindent
{\bf Visual Features.} For the visual frontend, we compute visual features $\tV \in \R^{M\times T\times D_v}$ with a 3D ConvNet \cite{Lecun98} on top of the synchronized video. We use a ``thick'' VGG-inspired block \cite{Simonyan15} exactly as in \cite{Braga20} and \cite{Makino19}, where the exact parameters of the stack of 5 layers of 3D convolutions can be found. During training, we have matched pairs of audio and a single corresponding face track, so $M$ is equal to the batch size $B$. During inference, $B = 1$ and $M$ is equal to the number of parallel face tracks in the video. For multi-person A/V ASR, we employ a single visual frontend, while for speaker selection and single-person A/V ASR we use two separately trained instances of this model. Each instance of the visual frontend has around 12 million parameters, so with two modules we  have considerably higher model capacity and cost in computation.

\begin{figure}[t]
  \centering
  \includegraphics[width=0.75\linewidth]{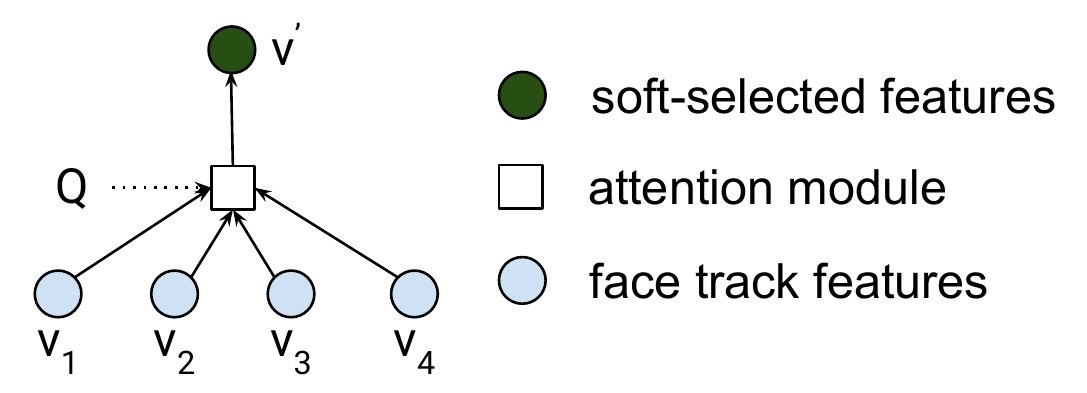}
  \caption{Face track soft-selection graph. At each step, the proposed soft-face selection mechanism has the structure of a gated attention graph.}
  \label{fig:graph}
\end{figure}

\subsection{Attention Mechanism for Soft Face Track Selection}
\label{sec:attention}
Visual features of competing faces are the slices $\tV_{m,:,:} \in \R^{T\times D_v}$ along the first dimension of the visual features tensor. We employ an attention \cite{BahdanauCB14,vaswani2017attention} module in order to soft-select the one matching the audio. Each entry $\tV_{m,t,:} \in \R^{D_v}$ can be seen as a memory cell we want to address. The attention queries $\tQ \in \R^{B\times T\times D_q}$ are computed with a 1D ConvNet of 5 layers on top of the acoustic features (consult~\cite{Braga20} for the exact parameters). The attention keys $\tK \in \R^{M\times T\times D_k}$, playing the role of addresses for the memory values in $\tV$, are also computed from the video.

Unlike our previous work \cite{Braga20}, where the attention keys and values are throughout assumed to be the same tensor (for us, computed from the video), here we consider the more general case where they are potentially distinct tensors extracted from separate submodules. For instance, on the speaker selection task the attention queries $\tQ$ and keys $\tK$ are trained separately from the values $\tV$, while on the multi-person A/V ASR model we make $\tV = \tK$ and train a single visual frontend. Figure~\ref{fig:graph} illustrates the attention graph.

We use a bilinear function with parameter matrix
$\tW \in \R^{D_q \times D_k}$ for the attention score, which can be expressed as
\begin{equation}
S_{btm} = Q_{btq}W_{qk}K_{mtk}, \quad \textrm{with} \quad \tS \in \R^{B\times T\times M}
\end{equation}
in Einstein summation notation. Note that during training $B = M$ since we have the same number of matched parallel audio and video tracks in a minibatch, and at inference time $B$ is $1$ and $M$ is the number of face tracks in the video, which can naturally be an arbitrary number since the attention is over a set.

In order to produce the attention weights, the attention scores are normalized with a softmax along their last dimension, {\it which corresponds to the batch dimension of size M of the keys and visual features}:
\begin{equation}
\label{eq:softmax}
\alpha_{btm} = \frac{e^{S_{btm}}}{\sum_{l} e^{S_{btl}}}, \quad \textrm{with} \quad \alpha \in \R^{B\times T \times M}.
\end{equation}

The attention weighted feature tensor $\tV'$ is then a weighted sum:
\begin{equation}
V'_{btk} = \alpha_{bti} V_{itk}, \quad \textrm{with} \quad \tV' \in \R^{B \times T \times D_v},
\label{eq:weighted_sum}
\end{equation}
and we end up with a new tensor of visual features, but now soft-selected among the competing visual feature vectors on the minibatch.

\subsection{Face Track Selection Model}
\label{sec:spk-selection}

A one-hot encoded face track selection function $z_t$ which picks at each timestep $t$ one among $m$ competing face tracks can be written as 
\begin{equation}
\label{eq:one_hot}
    z_t = \mathrm{one\_hot}\left(\argmax_m F_{tm}\right)\quad\in \R^{T\times M},
\end{equation}
where $F_{tm}$ is a scoring function which assigns, at each timestep, a scalar for each of the competing face tracks, with the highest score being assigned to the face track corresponding to the audio. $\argmax$ is not differentiable, but with the softmax approximation we can write
$z_{tm} \approx {e^{F_{tm}}}/{\sum_{l} e^{F_{tl}}}$ and Equation~\ref{eq:softmax} is simply a vectorized form of this equation. When using hard attention, the constraint is $||\alpha_{bt:}||_0=1$, while on soft-attention it is relaxed to $||\alpha_{bt:}||_1=1$.

This trick allows us to connect a speaker selection model to an A/V ASR model that can handle a single face track. When using the softmax approximation above, the whole system becomes differentiable and, thus, can be trained end-to-end with backpropagation.

On the other hand, during training we also have pairs of corresponding audio and video from a single speaking face, so we can train a model based on Equation~\ref{eq:softmax} directly with cross entropy for face track selection:
\begin{equation}
\label{eq:CE}
L = \frac{1}{BT}\sum_{b=1}^B\sum_{t=1}^T\sum_{m=1}^M-[b = m]\log \alpha_{btm}.
\end{equation}
Note that $\alpha$ only depends on the queries $\tQ$ and keys $\tK$, not on the values $\tV$ of the face tracks it is selecting. For our experiments we simply use a separate instance of the visual frontend, and for the query the same 1D ConvNet on top of the audio we described in the previous section.

During inference, at each timestep we compute the logits $\log \alpha_{1tm}$ for the $m$ competing face tracks on screen and select the track with the maximum value as the speaking face.

\begin{figure}[t]
  \centering
  \includegraphics[width=0.5\linewidth]{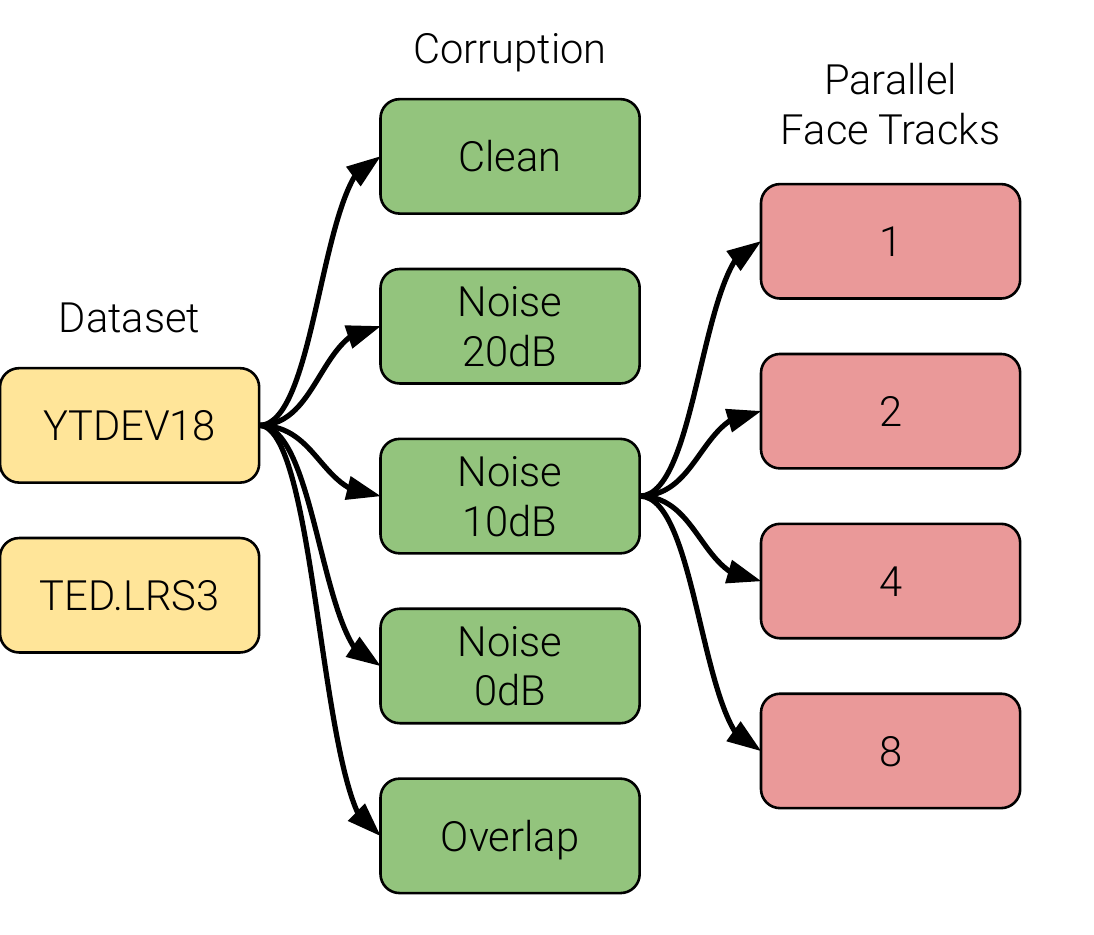}
  \caption{Augmented testing conditions starting from base test sets holding a single video face track matching the audio (see Section~\ref{sec:ms-datasets}).}
  \label{fig:test_sets}
\end{figure}

\subsection{RNN-T Encoder/Decoder Module for ASR}

For our experiments, we employ a standard RNN-T encoder-decoder as our ASR model \cite{graves2012sequence} \cite{graves2013speech} , with a stack of 5 BiLSTM of 512 units on each direction using layer normalization for the encoder, and 2 LSTM layers of 2048 units with character tokens for the decoder. For the input we fuse the acoustic $\tA$ and visual features $\tV'$ along the last dimension, yielding a combined feature tensor $\tF = [\tA; \tV'] \in \R^{B\times T \times (D_A + D_V)}$.

When training a multi-person model end-to-end, the visual features $\tV'$ is the attention weighted tensor from Equation~\ref{eq:weighted_sum}, and when training a single-person model it is simply the visual features tensor for the single track.

\begin{table}[!t]
%\small
\caption{{\bf Top-1 Face track selection accuracy at the frame level.} For each time step, we check if the model selection matches the ground truth track.}
\label{table:top-1-frame-acc}
\begin{center}
\begin{tabularx}{\linewidth}{cccccc}
\toprule
{\bf Dataset}                   & {\bf Noise}                & {\bf Tracks} &  {\bf SS} & {\bf [SS + A/V ASR]} \\
\midrule
\multirow{18}{*}{YTDEV18}	&		                     & 2            &  0.99                & 0.93 \\
                                &		                     & 4	        &  0.97                & 0.85 \\
                                &		                     & 8            &  0.95                & 0.77 \\
\cmidrule{2-5}
                                & \multirow{3}{*}{20dB}       & 2            &  0.99                 & 0.93 \\
                                &		                     & 4	        &  0.97                & 0.85 \\
                                &		                     & 8            &  0.95                & 0.77 \\
\cmidrule{2-5}
                                & \multirow{3}{*}{10dB}      & 2            &  0.98                & 0.91 \\
                                &		                     & 4	        &  0.96                & 0.83 \\
                                &		                     & 8            &  0.94                & 0.74 \\
\cmidrule{2-5}
                                & \multirow{3}{*}{0dB}      & 2            &  0.95                & 0.86 \\
                                &		                     & 4	        &  0.90                & 0.75 \\
                                &		                     & 8            &  0.83                & 0.63 \\
\cmidrule{2-5}
                                & \multirow{3}{*}{Overlap}   & 2            &  0.95                & 0.88 \\
                                &		                     & 4	        &  0.91                & 0.79 \\
                                &		                     & 8            &  0.87                & 0.70 \\
\midrule
\multirow{3}{*}{TED.LRS3}	    &		                     & 2            &  0.98                & 0.92 \\
                                &		                     & 4	        &  0.97                & 0.84 \\
                                &		                     & 8            &  0.94                & 0.74 \\
\bottomrule
\end{tabularx}
\end{center}
\vspace{-8mm}
\end{table}

\begin{table*}[!t]
%\singlespacing
\caption{{\bf Word error rates for the speech recognition task with multiple parallel video face tracks.} In each test example, the audio corresponds to a single face track. {\bf SS $\rightarrow$ A/V ASR} refers to a speaker selection model separately trained with CE loss coupled with an audio-visual ASR model that can accept a single face track assumed to correspond to the audio at training and inference time, and
{\bf Oracle SS $\rightarrow$ A/V ASR} indicates a similar setup but with oracle face track selection. {\bf [SS + A/V ASR]} refers to our proposed end-to-end model that can directly handle multiple video face tracks with an attention mechanism.}
\vspace{-3mm}
\aboverulesep = 0.2mm
\belowrulesep = 0.2mm
\label{table:wer-vs-num-speakers}
\begin{center}
\begin{tabularx}{\linewidth}{cccXcccc}
\toprule
%{\bf Dataset}                   & {\bf Noise}                & {\bf Tracks} & & {\bf AV\{0\}ASR}          & {\bf SS %$\rightarrow$ AV\{1\}ASR} & {\bf AV\{1,\}ASR} & {\bf SS* $\rightarrow$ AV\{1\}ASR}      \\
{\bf Dataset}                    & {\bf Noise}                & {\bf Tracks} & & {\bf Audio-only ASR}      & {\bf SS $\rightarrow$ A/V ASR}      & {\bf [SS + A/V ASR]} & {\bf Oracle SS $\rightarrow$ A/V ASR } \\
\midrule
\multirow{18}{*}{YTDEV18}	&		                     & 1            & &\multirow{4}{*}{16.5} & 15.5                 & 15.4              & \multirow{4}{*}{15.5}          \\
                                &		                     & 2	        & &	                     & 15.6                & 15.5             &                                \\
                                &		                     & 4            & &	                     & 15.7                & 15.7              &                                \\
                                &		                     & 8	        & &	                     & 15.8                & 15.8              &                                \\
\cmidrule{2-8}
                                & \multirow{4}{*}{20dB}      & 1            & &\multirow{4}{*}{16.9} & 15.7                & 15.6              & \multirow{4}{*}{15.7}          \\
                                &		                     & 2	        & &	                     & 15.7                & 15.7              &                                \\
                                &		                     & 4            & &	                     & 15.9                & 15.8              &                                \\
                                &		                     & 8	        & &	                     & 16.0                & 15.9              &                                \\
\cmidrule{2-8}
                                & \multirow{4}{*}{10dB}      & 1            & &\multirow{4}{*}{19.8} & 17.1                & 17.1              & \multirow{4}{*}{17.1}          \\
                                &		                     & 2	        & &	                     & 17.4               & 17.3              &                                \\
                                &		                     & 4            & &	                     & 17.7                & 17.6              &                                \\
                                &		                     & 8	        & &	                     & 18.1                & 17.9              &                                \\
\cmidrule{2-8}
                                & \multirow{4}{*}{0dB}       & 1            & &\multirow{4}{*}{42.8} &  26.6               & 26.7              & \multirow{4}{*}{26.6}          \\
                                &		                     & 2	        & &	                     & 28.8                & 27.9              &                                \\
                                &		                     & 4            & &	                     & 31.4                & 29.6              &                                \\
                                &		                     & 8	        & &	                     & 34.5                & 32.0              &                                \\
\cmidrule{2-8}
                                & \multirow{4}{*}{Overlap}   & 1            & &\multirow{4}{*}{34.3} & 30.4                &  29.7             & \multirow{4}{*}{30.4}          \\
                                &		                     & 2	        & &	                     & 31.3                & 30.7              &                                \\
                                &		                     & 4            & &	                     & 31.9                & 31.5              &                                \\
                                &		                     & 8	        & &	                     & 32.7                & 32.3               &                                \\
\midrule
\multirow{4}{*}{TED.LRS3}	    &		                     & 1            & &\multirow{4}{*}{3.8} & 3.7                & 3.6              & \multirow{4}{*}{3.8}          \\
                                &		                     & 2	        & &	                     & 3.7                & 3.7              &                                \\
                                &		                     & 4            & &	                     & 3.7                & 3.7              &                                \\
                                &		                     & 8	        & &	                     & 3.7                & 3.8              &                                \\
\bottomrule
\end{tabularx}
\end{center}
\label{table:track-selection-acc-av-asr}
\vspace{-8mm}
\end{table*}

\section{Datasets}
\label{sec:Datasets}

\

\noindent
{\bf Training.} For training, we use over 50k hours of transcribed short YouTube video segments extracted with the semi-supervised procedure originally proposed in \cite{Liao2013} and extended in \cite{Makino19,Shillingford_2019} to include video. We extract short segments where the force-aligned user uploaded transcription matches the transcriptions from a production quality ASR system. From these segments we then keep the ones in which the face tracks match the audio with high confidence.
 
\

\noindent
{\bf Evaluation.}
For the videos in both of our evaluation sets, we track the faces on screen and pick the segments with matching audio and video tracks with the same procedure used to extract the training data. Therefore, by design, the faces extracted from the video correspond with high probability to the speaker in the audio. We rely on two base datasets:
\begin{itemize}[leftmargin=*]
    \item {\it YTDEV18} \cite{Makino19}: Composed of 25 hours of manually transcribed YouTube videos, not overlapping with the training set, containing around 20k utterances.
    
    \item {\it LRS3-TED Talks} \cite{Afouras18d}: This is largest publicly available dataset for A/V ASR, so we evaluate on it as well for completeness. However, the dataset is not challenging enough as the performance quickly saturates with an audio only system and we don't observe significant gains when adding video. Videos are recorded by professionals, and the quality of both the audio and video are typically above the quality of a general video one would encounter in a realistic setting.
\end{itemize}

\subsection{Augmenting the Evaluation Sets with Noise and Parallel Video Tracks}
\label{sec:ms-datasets}
    First, in order to  measure the impact of the visual modality, we also evaluate on noisy conditions by adding babble noise randomly selected from the NoiseX dataset \cite{Varga1993AssessmentFA} at 0dB, 10dB and 20dB to each utterance. Moreover, we evaluate the effect of overlapping speech by randomly adding another utterance from the same dataset at the same level to the beginning and end of each utterance.
    
Secondly, in order to evaluate our model in the scenario where multiple face tracks are simultaneously visible in a video, we construct a new evaluation dataset as follows: On the single track evaluation sets described in the previous section, at time $t$ both the acoustic and visual features from the corresponding face are available. To build a dataset with $N$ parallel face tracks we start from the single track set, and for every pair of matched audio and face video track we randomly pick other $N-1$ face tracks from the same dataset. Therefore, during evaluation at each time step we have the acoustic features computed from the audio and $N$ candidate visual features, without knowing which one matches the audio. We generate separate datasets for $N = 1, 2, 4, 8$. The overall augmentation procedure is illustrated in figure~\ref{fig:test_sets}.

Note that these are potentially harder datasets than what we would normally encounter in practice since all faces are always speaking here, while it is an easier task to differentiate between a speaking and a non-speaking face video as we would encounter most of the time in a meeting scenario, for example. We adopt this evaluation protocol nonetheless and leave as future work to evaluate our model on more realistic datasets.

\section{Experiments}

\subsection{Baselines}

{\bf Audio-only ASR.} For the speech recognition task, we compare the performance of our audio-visual models against the performance of an equivalent audio-only baseline. This serves as a lower bound to the performance we need to achieve in order to show that the visual signal is helpful.

\medskip

\noindent
{\bf A/V ASR with oracle track selection (Oracle SS $\rightarrow$ A/V ASR).} This is a model trained with a single face track matching the audio. For the scenario with multiple speakers on screen this serves roughly as an upper bound to the performance we can hope to achieve with an audio-visual model, since this corresponds to an oracle selection of the correct video track that matches the audio.

\medskip

\noindent
{\bf  A/V Active Speaker Selection ({\bf SS}).} We use the weights of the attention module explicitly trained with cross-entropy loss, as outlined in Section~\ref{sec:spk-selection} to select the active speaker matching the audio. On the speech recognition task we use this model in conjunction with the single-track A/V ASR model for comparison with the multi-track end-to-end approach. On the active speaker selection task we use the speaker selection model by itself to measure the track selection accuracy.

\subsection{Training}

For training, we use the Adam~\cite{KingmaB14} optimizer with parameters $\beta_1 = 0.9$ and $\beta_2 = 0.98$. We adopt a learning rate schedule divided into 3 stages: The first stage is a linear warm-up to $0.001$ at $32{,}000$ steps, followed by keeping the learning rate constant at $0.001$ until $64{,}000$ steps, and then exponentially decaying it until $200{,}000$ steps, when we stop training. We clip the gradients to a maximum norm of $0.4$. Training is conducted on 128 TPU cores.

For all of our models, to augment the audio during training we use Multistyle TRaining (MTR) as described in \cite{Narayanan_2018}, which uses a room simulator to combine clean audio with a library of noise sources. This was not done in \cite{Braga20}, so now we are comparing against an even stronger audio-only baseline.

When training the multi-person model, we initialize the visual frontend from a model trained for single face tracks matching the audio, which we found to yield slightly better performance than training the whole model from scratch. This is not surprising as we are making the job of the visual module even harder when adding competing parallel tracks. By warm starting the visual frontend we make it easier for the video to ``catch-up'' with the audio from the start to some extent.

\subsection{Active Speaker Selection}

In order to measure how well the attention mechanism selects the correct face track and compare it to the face selection model explicitly trained for this task, we use the artificially constructed dataset with parallel video tracks described in Section~\ref{sec:Datasets}. At every 30ms (the duration between acoustic features) we select the face track with highest score (Equations~\ref{eq:one_hot} and \ref{eq:softmax}), and then count how often the first face track is picked, since the first track is always the ground truth speaker corresponding to the audio in our examples. Results are summarized in Table~\ref{table:top-1-frame-acc}, where we report the Top-1 accuracy of the attention mechanism ([SS + A/V ASR]) and of the model explicitly trained for speaker selection (SS). Even though the accuracy of the attention mechanism is consistently lower than of the speaker selection model, we can see it still gives surprisingly high accuracy on this challenging task even though it is never explicitly trained for it, just indirectly by association.

A second takeaway from Table~\ref{table:top-1-frame-acc} is that for the ASR task we are comparing our end-to-end model with a very strong baseline, since high accuracy in the baseline speaker selection model means that it is feeding the single track A/V ASR model the correct face track most of the time.

\vspace{-.5mm}
\subsection{Audio-Visual Speech Recognition}
\vspace{-1.mm}

The comparison of our proposed model ({\bf[SS + A/V ASR]}) with the baselines is summarized on Table~\ref{table:wer-vs-num-speakers}. First, we can verify that the model yields consistently lower WER than an audio only model, showing that the visual signal is indeed helping. For instance, with noise at 10dB and 0dB, even with 4 faces on screen we see a relative improvement of $12\%$ and $30\%$, respectively. On the other hand, when compared to the two steps system with hard attention between face selection and single face track A/V ASR (SS $\rightarrow$ A/V ASR), we see that our proposed model performs consistently better by a small margin. The main takeaway, however, is that we are employing a significantly smaller model (by around 12 million parameters) with lower computational costs and clearer formulation of the complete A/V ASR problem without losing performance. Lastly, when compared to the baseline A/V ASR model with oracle face selection we only show small degradation when adding more face tracks. For instance, at 10dB with 4 parallel tracks the degradation is about $3\%$ relative when compared to a single face track (and still 12\% relative better than the audio-only model, and about the same degradation when using an explicitly trained speaker selection model).

\vspace{-2.0mm}
\section{Conclusions}
\vspace{-1.5mm}

As it pertains to attention mechanisms, somewhat unique to our scenario in A/V ASR  is the fact that the visual input is secondary to the main task by a large margin to begin with (as lip reading is evidently a much harder task than speech recognition). To muddle things even further, we then added parallel video face tracks, making the signal even weaker. However, the model is still surprisingly able to make the association between the visual input and the output and pull useful information from the visual signal to complement the audio. One plausible alternate outcome for the training dynamics could have been to just deem the visual part of the encoder input as irrelevant to the task and only compress the information in the audio, which surprisingly does not happen. Our speaker selection experiment made it explicit that this is not taking place.

Some open questions that remain and we would like to explore in the future are: What happens when the video acts purely as a distractor, such as when the speaker is off-screen? Can we at the same time ignore the video in this case while still attending to it when it becomes helpful for the ASR task? Also, since we showed that we can clearly share a visual ``backbone'' for both speaker detection and ASR, so a natural question is whether we can get the best of both worlds with a multi-task training setup within this same attention framework.

\bibliographystyle{IEEEtran}

\bibliography{mybib}

\end{document}

%% file: math_commands.tex
\usepackage{amsmath,amsfonts,bm}

\def\eqref#1{equation~\ref{#1}}
\def\1{\bm{1}}

\DeclareMathAlphabet{\mathsfit}{\encodingdefault}{\sfdefault}{m}{sl}
\SetMathAlphabet{\mathsfit}{bold}{\encodingdefault}{\sfdefault}{bx}{n}
\newcommand{\tens}[1]{\bm{\mathsfit{#1}}}
\def\tA{{\tens{A}}}

\def\tF{{\tens{F}}}

\def\tK{{\tens{K}}}

\def\tQ{{\tens{Q}}}

\def\tS{{\tens{S}}}

\def\tV{{\tens{V}}}
\def\tW{{\tens{W}}}

\newcommand{\R}{\mathbb{R}}

\DeclareMathOperator*{\argmax}{arg\,max}

%% file: main.bbl
\begin{thebibliography}{10}
\providecommand{\url}[1]{#1}
\csname url@samestyle\endcsname
\providecommand{\newblock}{\relax}
\providecommand{\bibinfo}[2]{#2}
\providecommand{\BIBentrySTDinterwordspacing}{\spaceskip=0pt\relax}
\providecommand{\BIBentryALTinterwordstretchfactor}{4}
\providecommand{\BIBentryALTinterwordspacing}{\spaceskip=\fontdimen2\font plus
\BIBentryALTinterwordstretchfactor\fontdimen3\font minus
  \fontdimen4\font\relax}
\providecommand{\BIBforeignlanguage}[2]{{%
\expandafter\ifx\csname l@#1\endcsname\relax
\typeout{** WARNING: IEEEtran.bst: No hyphenation pattern has been}%
\typeout{** loaded for the language `#1'. Using the pattern for}%
\typeout{** the default language instead.}%
\else
\language=\csname l@#1\endcsname
\fi
#2}}
\providecommand{\BIBdecl}{\relax}
\BIBdecl

\bibitem{Makino19}
T.~Makino, H.~Liao, Y.~Assael, B.~Shillingford, B.~Garcia, O.~Braga, and
  O.~Siohan, ``Recurrent neural network transducer for audio-visual speech
  recognition,'' \emph{ASRU}, 2019.

\bibitem{Afouras_2018}
T.~Afouras, J.~S. Chung, A.~Senior, O.~Vinyals, and A.~Zisserman, ``Deep
  audio-visual speech recognition,'' \emph{IEEE Transactions on Pattern
  Analysis and Machine Intelligence}, p. 1–1, 2018.

\bibitem{Chung_2017}
\BIBentryALTinterwordspacing
J.~S. Chung, A.~Senior, O.~Vinyals, and A.~Zisserman, ``Lip reading sentences
  in the wild,'' \emph{2017 IEEE Conference on Computer Vision and Pattern
  Recognition (CVPR)}, Jul 2017. [Online]. Available:
  \url{http://dx.doi.org/10.1109/cvpr.2017.367}
\BIBentrySTDinterwordspacing

\bibitem{Potamianos2003}
G.~{Potamianos}, C.~{Neti}, G.~{Gravier}, A.~{Garg}, and A.~W. {Senior},
  ``Recent advances in the automatic recognition of audiovisual speech,''
  \emph{Proceedings of the IEEE}, vol.~91, no.~9, pp. 1306--1326, 2003.

\bibitem{Saenko2006}
K.~{Saenko} and K.~{Livescu}, ``An asynchronous dbn for audio-visual speech
  recognition,'' in \emph{2006 IEEE Spoken Language Technology Workshop}, 2006,
  pp. 154--157.

\bibitem{Harte2015}
N.~Harte and E.~Gillen, ``Tcd-timit: An audio-visual corpus of continuous
  speech,'' \emph{Multimedia, IEEE Transactions on}, vol.~17, pp. 603--615, 05
  2015.

\bibitem{ChungSyncNet_2017}
J.~S. Chung and A.~Zisserman, ``Out of time: Automated lip sync in the wild,''
  in \emph{Workshop on Multi-view Lip-reading, ACCV, 2016}, 03 2017, pp.
  251--263.

\bibitem{ChungSyncnet2_2019}
\BIBentryALTinterwordspacing
S.-W. Chung, J.~S. Chung, and H.-G. Kang, ``Perfect match: Improved cross-modal
  embeddings for audio-visual synchronisation,'' \emph{ICASSP 2019 - 2019 IEEE
  International Conference on Acoustics, Speech and Signal Processing
  (ICASSP)}, May 2019. [Online]. Available:
  \url{http://dx.doi.org/10.1109/icassp.2019.8682524}
\BIBentrySTDinterwordspacing

\bibitem{Braga20}
O.~{Braga}, T.~{Makino}, O.~{Siohan}, and H.~{Liao}, ``End-to-end multi-person
  audio/visual automatic speech recognition,'' in \emph{ICASSP 2020 - 2020 IEEE
  International Conference on Acoustics, Speech and Signal Processing
  (ICASSP)}, 2020, pp. 6994--6998.

\bibitem{GoogleAIPrinciples}
{Artificial Intelligence at Google: Our Principles}.
  \url{https://ai.google/principles/}.

\bibitem{Lecun98}
Y.~Lecun, L.~Bottou, Y.~Bengio, and P.~Haffner, ``Gradient-based learning
  applied to document recognition,'' in \emph{Proceedings of the IEEE}, 1998,
  pp. 2278--2324.

\bibitem{Simonyan15}
K.~Simonyan and A.~Zisserman, ``Very deep convolutional networks for
  large-scale image recognition,'' in \emph{International Conference on
  Learning Representations}, 2015.

\bibitem{BahdanauCB14}
\BIBentryALTinterwordspacing
D.~Bahdanau, K.~Cho, and Y.~Bengio, ``Neural machine translation by jointly
  learning to align and translate,'' in \emph{3rd International Conference on
  Learning Representations, {ICLR} 2015, San Diego, CA, USA, May 7-9, 2015,
  Conference Track Proceedings}, 2015. [Online]. Available:
  \url{http://arxiv.org/abs/1409.0473}
\BIBentrySTDinterwordspacing

\bibitem{vaswani2017attention}
A.~Vaswani, N.~Shazeer, N.~Parmar, J.~Uszkoreit, L.~Jones, A.~N. Gomez,
  {\L}.~Kaiser, and I.~Polosukhin, ``Attention is all you need,'' in
  \emph{Advances in neural information processing systems}, 2017, pp.
  5998--6008.

\bibitem{graves2012sequence}
A.~Graves, ``Sequence transduction with recurrent neural networks,''
  \emph{ICML}, 2012.

\bibitem{graves2013speech}
A.~Graves, A.-r. Mohamed, and G.~Hinton, ``Speech recognition with deep
  recurrent neural networks,'' in \emph{2013 IEEE international conference on
  acoustics, speech and signal processing}.\hskip 1em plus 0.5em minus
  0.4em\relax IEEE, 2013, pp. 6645--6649.

\bibitem{Liao2013}
H.~{Liao}, E.~{McDermott}, and A.~{Senior}, ``Large scale deep neural network
  acoustic modeling with semi-supervised training data for youtube video
  transcription,'' in \emph{2013 IEEE Workshop on Automatic Speech Recognition
  and Understanding}, 2013, pp. 368--373.

\bibitem{Shillingford_2019}
\BIBentryALTinterwordspacing
B.~Shillingford, Y.~Assael, M.~W. Hoffman, T.~Paine, C.~Hughes, U.~Prabhu,
  H.~Liao, H.~Sak, K.~Rao, L.~Bennett, and et~al., ``Large-scale visual speech
  recognition,'' \emph{Interspeech 2019}, Sep 2019. [Online]. Available:
  \url{http://dx.doi.org/10.21437/interspeech.2019-1669}
\BIBentrySTDinterwordspacing

\bibitem{Afouras18d}
T.~Afouras, J.~S. Chung, and A.~Zisserman, ``Lrs3-ted: a large-scale dataset
  for visual speech recognition,'' in \emph{arXiv preprint arXiv:1809.00496},
  2018.

\bibitem{Varga1993AssessmentFA}
A.~Varga and H.~J.~M. Steeneken, ``Assessment for automatic speech recognition:
  Ii. noisex-92: A database and an experiment to study the effect of additive
  noise on speech recognition systems,'' \emph{Speech Communication}, vol.~12,
  pp. 247--251, 1993.

\bibitem{KingmaB14}
\BIBentryALTinterwordspacing
D.~P. Kingma and J.~Ba, ``Adam: {A} method for stochastic optimization,'' in
  \emph{3rd International Conference on Learning Representations, {ICLR} 2015,
  San Diego, CA, USA, May 7-9, 2015, Conference Track Proceedings}, Y.~Bengio
  and Y.~LeCun, Eds., 2015. [Online]. Available:
  \url{http://arxiv.org/abs/1412.6980}
\BIBentrySTDinterwordspacing

\bibitem{Narayanan_2018}
\BIBentryALTinterwordspacing
A.~Narayanan, A.~Misra, K.~C. Sim, G.~Pundak, A.~Tripathi, M.~Elfeky,
  P.~Haghani, T.~Strohman, and M.~Bacchiani, ``Toward domain-invariant speech
  recognition via large scale training,'' \emph{2018 IEEE Spoken Language
  Technology Workshop (SLT)}, Dec 2018. [Online]. Available:
  \url{http://dx.doi.org/10.1109/SLT.2018.8639610}
\BIBentrySTDinterwordspacing

\end{thebibliography}
